\documentclass{aa} 

\usepackage{graphicx}
\usepackage{txfonts}
\usepackage[
 breaklinks,
 colorlinks=true,
 linkcolor=blue,
 citecolor=blue,
 urlcolor=blue
]{hyperref}

\usepackage{multirow}

\begin{document}

\title{Impact of charge transfer inefficiency on transit light curves} 

\subtitle{A correction strategy for PLATO}

  \author{S. Mishra\inst{1}
    \and R. Samadi\inst{1}
    \and D. Bérard\inst{1}
    }

  \institute{LIRA, Observatoire de Paris, Université PSL, Sorbonne Université, Université Paris Cité, CY Cergy Paris Université, CNRS,  92190 Meudon, France}
  \mail{shaunak.mishra@obspm.fr}
\date{Received 29 August 2025 / Accepted <to be set by the editor>}

\abstract
 {PLATO is designed to detect Earth-sized exoplanets orbiting solar-type stars and to measure their radii (relative to the star radii)  with an accuracy better than \(2\%\) via the transit method. Charge transfer inefficiency (CTI), a by-product of radiation damage to charge-coupled devices (CCDs), can jeopardise this accuracy constraint and therefore must be corrected to reach scientific requirements.}
 {We assessed and quantified the impact of CTI on transit depth measurements. Our objective was to demonstrate the need for CTI correction and to develop a correction strategy that restores CTI-biased transit depths with an acceptable residual within the accuracy budget.}
 {Using a calibration dataset generated with PLATOSim to simulate a realistic stellar field, we modelled the parallel overscan signal as the sum of exponential decays and used least-squares fitting to infer the number of trap species and initial estimates for the release times (\(\tau_{\mathrm{r, k}}\)). Smearing was then modelled with an exponential-plus-constant function and removed on a column-wise basis. We modelled the spatial variation in the trap density with a quadratic polynomial function of the radial distance from the centre of the focal plane. The polynomial coefficients (\(a_{p,k}\)) of this model, the well-fill power index (\(\beta\)), and the release times (\(\tau_{\mathrm{r, k}}\)) were subsequently adjusted via an iterative application of the extended pixel edge response method combined with a CTI correction algorithm. This yielded the final calibration model that underpins our correction strategy.}
 {In the worst-case scenario (8-year mission, high CTI impact zone), we found that CTI induced a bias of approximately \(4\%\) in the measured transit depth. The polynomial coefficients from our trap density model were then used to correct the CTI-affected transit depths. Our correction reduced the bias to a residual of \(0.06\%\), which is comfortably within PLATO's accuracy requirements.}
 {We quantified the CTI-induced bias in transit depth measurements and implemented a calibration strategy that incorporates spatial variations in trap density. From the calibrated parameters, we derived a correction scheme that brought the photometric measurements within PLATO's noise budget, ensuring that the mission's precision requirements are met.}


  \maketitle
  \nolinenumbers

\section{Introduction}
\label{sec:introduction}

The PLAnetary Transits and Oscillations of stars (PLATO) is the European Space Agency's M3 mission and is scheduled for launch in late 2026. The goal of the PLATO mission is to detect and describe exoplanets and conduct asteroseismological studies of host and observed stars. Specifically, it will identify Earth-sized exoplanets (radius < \(2R_{Earth}\)) orbiting F5-K7 dwarf and sub-giant stars (magnitude $\lesssim~11$), with a particular focus on terrestrial planets located in the habitable zone of solar-type stars. Thus, it will offer critical insights into planet formation and the existence of comparable solar systems and potentially habitable planets \citep{Rauer_2025}.

PLATO will conduct high-precision photometric observations to identify these planets via the transit method. According to PLATO's accuracy requirements, the planet-to-star radius ratio for an Earth-sized planet orbiting a G0V star of \(m_{V} = 10\) (goal \(m_{V} = 11\)) must be determined with an accuracy better than \(2\%\) \citep{PLATO2019SciRD}.

This noise budget has to be shared among different sources of systematic error. However, as we show in the present work, meeting these precision requirements will be challenging because the space environment can damage the charge-coupled devices (CCDs) that enable the observations. The payload's exposure to radiation will induce defects in the CCDs, creating charge-trapping sites that intermittently capture and release electrons. This deferred charge release causes flux loss outside the aperture mask and degrades photometric observations. This phenomenon, known as charge transfer inefficiency (CTI), affects the quality of light curves used to extract transit depths and, consequently, introduces measurement errors. While PLATO's noise budget does not explicitly allocate CTI, the consortium's internal reference value for sky background noise (\(0.4\%\)) serves as the benchmark threshold for CTI mitigation.

Modelling and assessing the impact of CTI on astrophysical applications has been carried out for missions such as \textit{Euclid} and the FLuorescence EXplorer (FLEX; scheduled for launch in 2026; \citealt{Israel_2015, Bernard_2021}). For \textit{Euclid}'s CCD detectors, a CTI correction pipeline called ArCTIC, built on the Pyxel framework \citep{Matej_2022}, has been developed \citep{Kelman_2022, Kelman_2024}. While these existing methods rely on robust frameworks, they do not account for the spatial variation in trap density across the detector for each species. Ignoring such spatial variations may bias correction accuracy in real missions. Our work incorporates the spatially varying nature of trap density, which is anticipated for PLATO. To our knowledge, this is the first quantitative demonstration of CTI impacting exoplanet transit measurements. Although our study focuses on PLATO, the calibration and correction strategy presented here is broadly applicable to any detector system affected by CTI, providing a general framework for restoring CTI-impacted transit depths and preserving photometric accuracy.

To demonstrate the need for CTI mitigation, we quantified the CTI-induced bias in the transit depth measurement of an Earth-sized exoplanet transiting a solar-type star in the context of PLATO's accuracy requirements (Sect.~\ref{sec:cti_impact_transit_depth}).
We present our CTI calibration strategy in Sect.~\ref{sec:cti_calibration_strategy}. CTI is modelled using the charge density model (CDM; \citealt{Short_2013,Massey_2014}), which is based on the Shockley--Read--Hall theory \citep{Shockley_1952,Hall_1952,Hardy_1998}. The CDM depends on the mean trap density (\(n_\mathrm{t}\)), release time (\(\tau_\mathrm{r}\)), well-fill power index (\(\beta\)), capture cross-section (\(\sigma\)), and operating temperature. To extend CDM beyond its uniform mean trap density assumption, we adopted a trap density distribution, incorporating an axisymmetric radial dependence inferred from the radiation map in \citet{ohb2024radiation}. The calibration proceeds in three stages: (1) initial iterative estimates of trap release times, (2) a non-iterative smearing correction via least squares, and (3) an iterative fit of the polynomial coefficients of the trap density model, well-fill power-law index (\(\beta\)), and the release times via an algorithm involving the extended pixel edge response (EPER) method combined with the \citet{Massey_2014} CTI correction algorithm.

Section~\ref{sec:cti_correction_strategy} outlines our CTI correction strategy, which is adapted from the iterative inversion algorithm of \citet{Massey_2014} and relies on the CTI parameters obtained from our calibration procedure. We evaluate the performance of this correction scheme, demonstrating its ability to recover CTI-affected transit depths toward their true values. Finally, Sect.~\ref{sec:conclusions_discussion} summarises our findings, highlights the limitations of the calibration-correction framework, and discusses the assumptions underlying our approach.

\section{CTI and its impact on transit depth}
\label{sec:cti_impact_transit_depth}

\subsection{The PLATO instrument and the photometric measurements}
\label{subsec:plato_photometric_requirements}

The PLATO payload consists of 26 cameras: 24 `normal' cameras operating in white light for high-precision photometry, and 2 `fast' cameras providing colour information in the blue (505--700 nm) and red (665--1000 nm) bands \citep{Marchiori_2019}. Each camera hosts four CCD detectors at its focal plane. 
About \(80\%\) of PLATO targets will have on-board photometry extracted using fixed aperture masks. Given the limited on-board CPU resources, CTI correction at the pixel level is not feasible in this mode. 
For the remaining targets, which largely constitute the primary sample (S1), small \(6 \times 6\) pixel `imagettes' will be transmitted to ground at a cadence of \(25\,\)s for the normal cameras and \(2.5\,\)s for the fast cameras. The photometry of these targets will be extracted on ground. In this work, we therefore designed and validated a CTI correction strategy specifically for the imagettes.

\subsection{Basic characteristics of CTI}
\label{subsec:basic_characteristics_cti}
\begin{figure}
  \centering
  \includegraphics[width=\columnwidth]{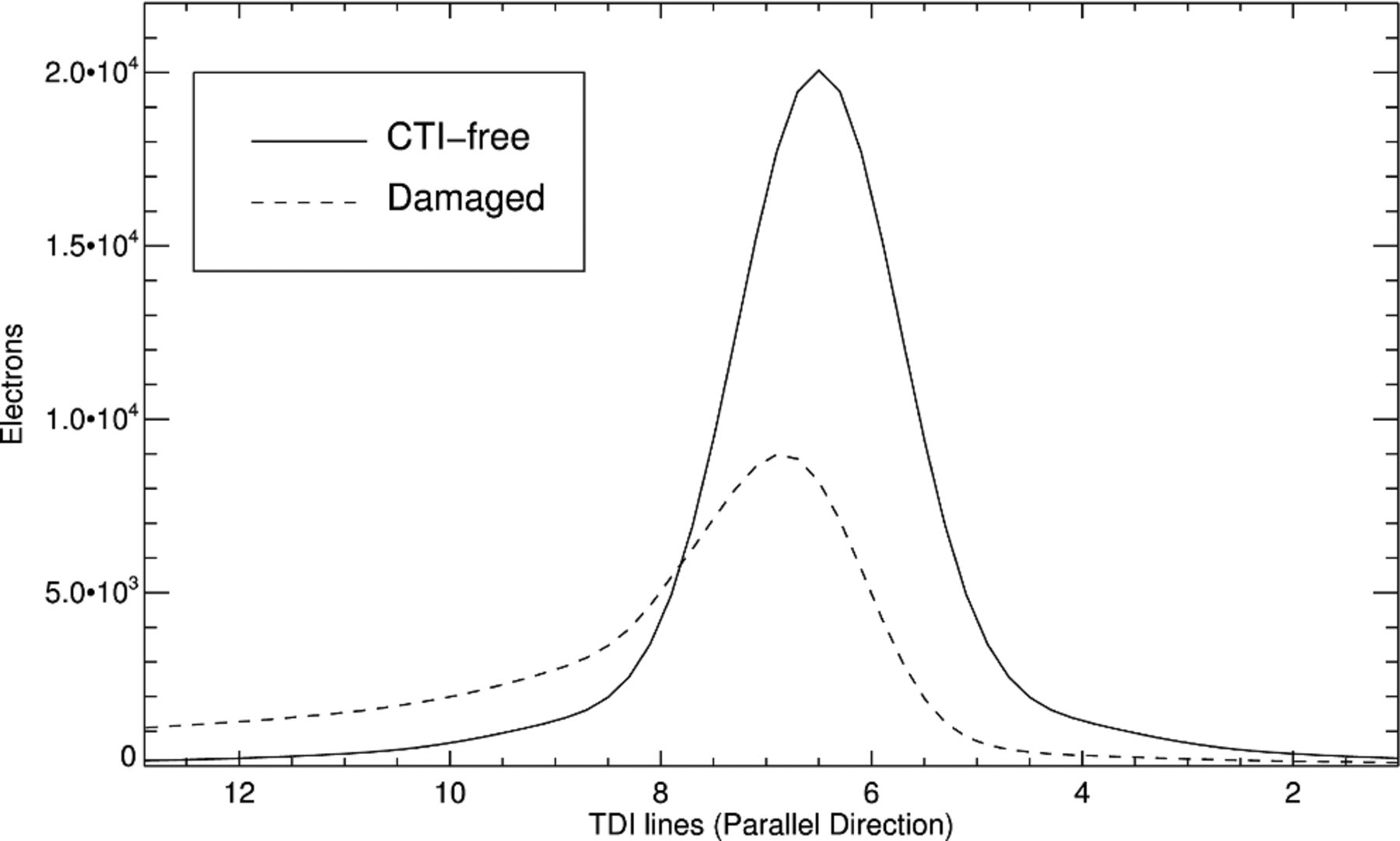}
  \caption{Effect of CTI on the recorded signal in the parallel transfer direction (readout proceeds from left to right). The solid curve shows a CTI-free signal profile, while the dashed curve illustrates the same profile after CTI degradation. Deferred charge release reduces the peak amplitude and produces a trailing towards increasing time delay integration (TDI) lines (earlier pixels in the transfer path). Credit: \citet{Short_2013}.}
  \label{fig:cti_effect}
\end{figure}

Charge transfer inefficiency is a by-product of prolonged exposure of CCDs to radiation in space, from sources such as cosmic rays and solar particles. Radiation-induced defects create trapping sites in the silicon lattice that capture and later release charges during transfer stage in the CCDs. During its in-flight operation, the instruments' prolonged exposure to radiation will increase the number of charge trapping sites and thus, intensify CTI. This will impact the measured signal, causing charge trailing along the CCD columns (Fig.~\ref{fig:cti_effect}). This delayed release of electrons causes redistribution of flux from the source, leading to distorted image and potential photometric biases. 

Charge transfer inefficiency is a serious concern for PLATO, which relies on photometric time series to detect transits and identify exoplanets, as the charge-release timescale is of the order of \(10^{-4}\,\mathrm{s}\) for the fastest traps and \(10^{-1}\,\mathrm{s}\) for the slowest traps. These values are --~for the majority of the trap species~-- significantly longer (particularly for the slowest traps) than the transfer cadence of about \(900~\mu\mathrm{s}\), posing a risk to the robustness of the observations.

\subsection{CTI modelling}
\label{subsec:cti_induction_process}
Various strategies exist to simulate parallel CTI effects, differing mainly in their treatment of trap occupancy and integration. In this work, we used the CDM, which models charge capture and release probabilistically and governs the interaction between traps and the electron cloud \citep[see e.g.][and reference therein]{Short_2013,Massey_2014}. Specifically, following the formalism of \citet{Short_2013}, the probability of trap capture depends on the local electron density and is controlled by the parameter \(\beta\), which defines how the electron cloud volume scales with the number of electrons. A larger electron cloud increases trap accessibility, and this dependence is expressed as
\begin{equation}
\frac{V_{\mathrm{c}}}{V_{\mathrm{g}}}=\left(\frac{N_{\mathrm{e}}}{\mathrm{FWC}}\right)^{\beta}\, ,
 \label{eq:eq_for_beta}
\end{equation}

\noindent where $V_{\mathrm{c}}$ is volume of the charge cloud in a pixel; $V_{\mathrm{g}}$ is maximum geometrical volume a pixel can hold; $N_{\mathrm{e}}$ is the number of electrons in the pixel; $\mathrm{FWC}$ is full-well capacity of the pixel (the maximum electron load); and $\beta$ is the well-fill power-law index ($0\le\beta\le1$ in practice).

Charge transfer inefficiency can be modelled using the \texttt{run\_cdm\_parallel} routine in the Pyxel\footnote{\url{https://esa.gitlab.io/pyxel/}} framework \citep{Matej_2022}. Based on the \citet{Short_2013} formulation, this routine requires as inputs: the well-fill power index (\(\beta\)), the trap release times (\(\tau_\mathrm{r}\)), the trap densities (\(n_\mathrm{t}\)), and capture cross-sections (\(\sigma\)), which control the probability of interactions between charges and traps for each involved trap species. \citet{Short_2013} formalism assumes uniform trap densities. In practice, this is not the case as we describe in Sect.~\ref{subsec:radiation_analysis}. However, as explained below, it is possible to generalise this formalism in order to account for an arbitrary distribution of traps. 

For a uniform trap distribution, the number of traps (for a given trap species~\(k\)) encountered by a charge packet during transfer from row index~\(i\) to the register is given by \(n_{t,k}\, y\), where \(n_{t,k}\) is the (constant) trap density and \(y = i + 1\) denotes the number of transfers (i.e. the distance from row index~\(i\) to the register).

When the trap distribution varies spatially, the term \(n_{t,k}\, y\) is replaced by \(cnt_k[i,j]\), which represents the cumulative number of traps crossed by a charge packet during its transfer from row \(i\) down to the register, for a given column \(j\). We extended the CDM accordingly by introducing \texttt{run\_cdm\_parallel\_cumul} (see the Python implementation in \href{https://github.com/mishrashaunak/plato-cti-correction}{plato-cti-correction}). Unlike the original routine, which assumes a constant trap density per pixel, our adaptation accounts for the cumulative trap content encountered during vertical transfer. This captures the integrated effect of spatial variations in the trap distribution.

Trap species parameters from \citet{Prodhomme_2016}, including the well-fill power index (\(\beta\)), release times (\(\tau_{r,k}\)), capture cross-sections (\(\sigma_k\)), and reference mean trap densities (\(n_{t,k}^{\mathrm{ref}}\)), are listed in Table~\ref{table:table_1}. In this study, we adopted these reference values: \(\beta = 0.37\), species-specific \(\sigma_k\) and \(\tau_{r,k}\). The trap densities (\(n_{t,k}\)) are subsequently employed to construct the cumulative trap distribution, as described in Sect.~\ref{subsec:radiation_analysis}.

\begin{table}[ht]
 \caption{Reference trap species parameters.}
 \label{table:table_1}
 \centering
 \renewcommand{\arraystretch}{1.25}
 \setlength{\tabcolsep}{4pt}
 \begin{tabular}{ccccc}
  \hline\hline
  Species & Type & $\tau_\mathrm{r, k}$ & $\sigma_k$ & $n_\mathrm{t, k}^{\mathrm{ref}}$ \\
    &  & (s)         & ($\mathrm{m}^{-2}$) & (per pixel) \\
  \hline
   1 & Fast      & $2.37\times10^{-4}$ & $2.46\times10^{-20}$ & 9.80 \\
   2 & Intermediate & $2.03\times10^{-3}$ & $7.05\times10^{-23}$ & 3.31 \\
   3 & Intermediate & $2.43\times10^{-2}$ & $1.74\times10^{-22}$ & 1.56 \\
   4 & Slow      & $1.40\times10^{-1}$ & $2.45\times10^{-23}$ & 13.24 \\
  \hline
 \end{tabular}
 \tablefoot{Parameters are from \citet{Prodhomme_2016} for a fluence of \(13\times10^{9}\,\mathrm{cm}^{-2}\). The well-fill power index is fixed at $\beta = 0.37$.}
\end{table}

\subsection{Radiation analysis}
\label{subsec:radiation_analysis}
\begin{figure}
  \centering
  \includegraphics[width=\columnwidth]{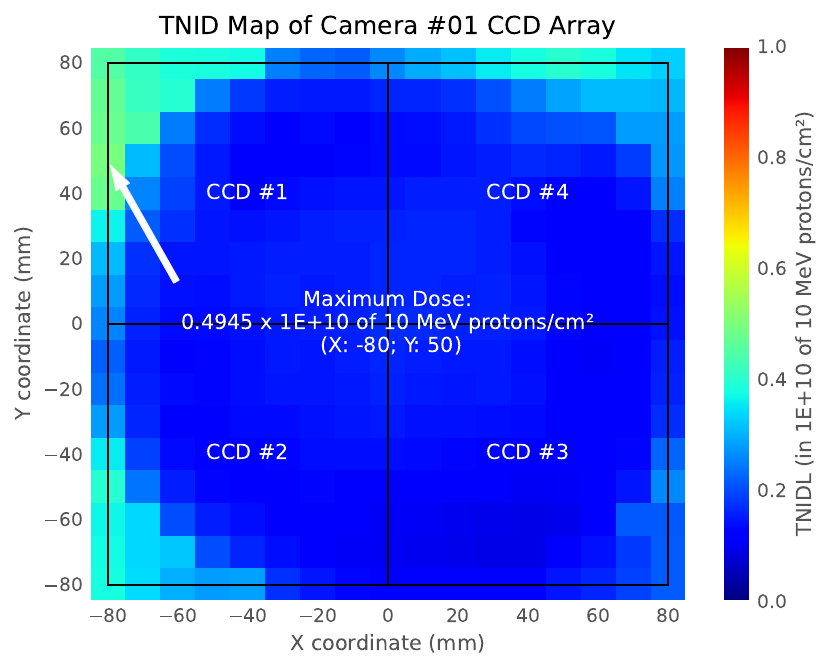}
\caption{Radiation map of Camera~1, composed of four CCDs, showing the radial dependence of the trap density. The colour bar indicates the TNID level (trap density) to which each pixel is exposed after 6.5~years. Adapted from \citet{ohb2024radiation}.}
\label{fig:og_radiation_map}
\end{figure}

The PLATO radiation analysis by \citet{ohb2024radiation} reports the total non-ionising dose (TNID) for Camera~1, which comprises four CCDs, after 6.5 years of exposure. The radiation map, displayed in Fig.~\ref{fig:og_radiation_map}, indicates that TNID---and therefore trap density---increases approximately with the square of the radial distance from the centre, implying a first-order axisymmetric distribution across the CCD.

To reproduce the radiation map shown in Fig.~\ref{fig:og_radiation_map}, we generated a grid of coordinates across the CCD and computed the normalised radial distance (\(R_{\mathrm{norm}}\)) using

\begin{equation}
    R_{\mathrm{norm}} = \frac{1}{\sqrt{2}} 
    \sqrt {    
    \left( \frac{j - x_c}{4510} \right)^2 + 
    \left( \frac{i - y_c}{4510} \right)^2
       }\, ,
\label{eq:Rnorm}
\end{equation}
where \((x_c, y_c)\) denotes the principal point (intersection of the optical axis with the focal plane) in the CCD reference frame. 

\begin{figure}
  \centering
  \includegraphics[width=\columnwidth]{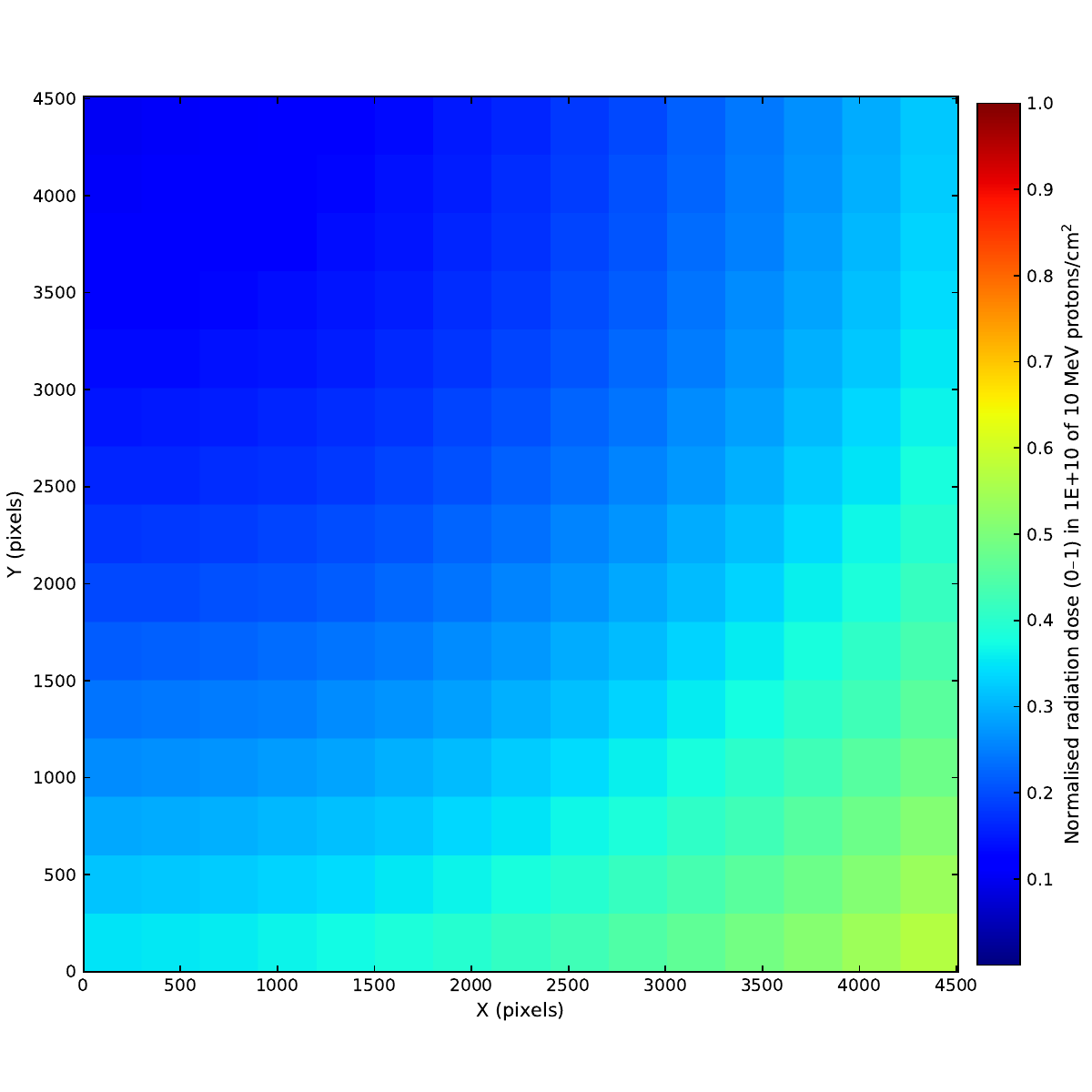}
\caption{Radiation map generated using the radial dependence model (Eq.~\ref{eq:radial_relationship}) for a 6.5-year TNID exposure. The CCD is oriented with the readout register at the bottom (\(y = 0\)). The simulated distribution shows strong agreement with the reference radiation map for CCD~\#1 (Fig.~\ref{fig:og_radiation_map}), adapted from \citet{ohb2024radiation}, under identical TNID conditions.}
 \label{fig:our_radiation_map}
\end{figure}

The TNID distribution is approximately modelled as a quadratic function of the normalised radial distance (\(R_{\mathrm{norm}}\)):
\begin{equation}
 D(R_{\mathrm{norm}}) = D_{\mathrm{max}}
 \left[
   \frac{D_{\mathrm{min}}}{D_{\mathrm{max}}}
   + \left(
     1 - \frac{D_{\mathrm{min}}}{D_{\mathrm{max}}}
    \right) R_{\mathrm{norm}}^{2}
 \right]\, ,
 \label{eq:radial_relationship}
\end{equation}
varying between a minimum dose ($D_{\mathrm{min}}$) at the radial origin and a maximum dose ($D_{\mathrm{max}}$) at the edge. Figure~\ref{fig:our_radiation_map} compares the radiation map derived from Eq.~\ref{eq:radial_relationship} with the reference distribution in Fig.~\ref{fig:og_radiation_map}, showing close agreement for CCD~\#1 under identical TNID conditions.
The residual departure from our axisymmetric model is discussed in Sect.~\ref{sec:conclusions_discussion}.

For a 6.5-year mission, the maximum TNID reported by \citet{ohb2024radiation} corresponds to a total fluence of \(5\times10^{9}\,\mathrm{cm}^{-2}\). Assuming a constant radiation rate, the annual fluence is computed via linear scaling:

\begin{equation}
  \Phi_{\mathrm{year}} = \frac{5\times10^{9}}{6.5}\;\mathrm{cm}^{-2}\,\mathrm{yr}^{-1}\, .
  \label{eq:annual_fluence}
\end{equation}

The TNID for 4-year (nominal) and 8-year (extended) duration is then obtained by multiplying \(\Phi_{\mathrm{year}}\) by the respective mission duration. The reference mean trap densities (\(n_{t,k}^{\mathrm{ref}}\)) in Table~\ref{table:table_1} correspond to a TNID of \(13\times10^{9}\,\mathrm{cm}^{-2}\) \citep{Prodhomme_2016}. To account for mission-specific doses, these values are scaled linearly as
\begin{equation}
 n_{t,k}^{\mathrm{scaled}} =
  \left(
   \frac{\Phi_{\mathrm{year}}\times \mathrm{T}}
      {13\times10^{9}}
  \right)
  n_{t,k}^{\mathrm{ref}}\, ,
 \label{eq:nt_prodhomme_scaling}
\end{equation}
where T is the desired mission duration in years. The scaled trap densities are then distributed across the CCD according to an axisymmetric radial model, that is, 
\begin{equation}
{n}_{t,k}[i, j] = D (R_{\mathrm{norm}}[i,j])  \, n_{t,k}^{\mathrm{scaled}}\, ,
\label{eq:nt_true}
\end{equation}
where \(D\) is given in Eq.~\ref{eq:radial_relationship}. 

A cumulative sum along the row axis yields a three-dimensional array representing the \texttt{cnt} parameter:\begin{equation}
  {cnt}_{k}[i, j] = \sum_{i'=0}^{i}{n}_{t,k}[i', j]\, .
  \label{eq:cnt_true}
\end{equation}
This quantity serves as input for the CTI simulations in subsequent steps.

\subsection{Bias on transit depth measurements}
\label{subsec:bias_transit_depth_measurements}
In the context of the axisymmetric radial dependence (Eq.~\ref{eq:nt_true}), we modelled CTI using the routine \texttt{cdm.run\_cdm\_parallel\_cumul} (see Sect.~\ref{subsec:cti_induction_process}), which takes as input the cumulative trap distribution defined in Eq.~\ref{eq:cnt_true}.
Based on this CTI model, with trap densities scaled for the mission duration and CCD position, we simulated the resulting bias in planetary transit measurements. 

To assess the impact of CTI on the transit method, we simulated representative light curves for both Earth- and Jupiter-sized planets transiting a solar-type star. The intrinsic stellar transit signal is generated using the PLATO Solar-like Light-curve Simulator\footnote{\url{https://sites.lesia.obspm.fr/psls/}} \citep{Samadi_2019}, which is based on the analytical model by \citet{Mandel_2002} and its Python implementation by Ian Crossfield (MIT)\footnote{\url{http://www.astro.ucla.edu/~ianc/}}.

The flux variations over time are interpolated onto a grid of 2000 time steps. The stellar image is modelled using a Gaussian point spread function (PSF) with a characteristic width of \(0.6\)~pixel. The total flux is scaled to match the expected value for a given stellar magnitude, following the flux--magnitude relation from \citet{Marchiori_2019}. Finally, this flux distribution is modulated by the intrinsic transit light curve.

We added a constant background contribution from zodiacal light of \(45~\mathrm{e^-\,s^{-1}\,pixel^{-1}}\) \citep{Marchiori_2019}. Given the integration time of \(21\)s for the PLATO CCD, the net background value is \(\approx1000~\mathrm{e^-\,pixel^{-1}}\). A higher background fills traps that would otherwise capture flux from the target star, thereby reducing the CTI effect. The aperture size is determined using our optimal aperture mask routine, following the approach of \citet{Marchiori_2019}. Finally, from the simulated light curve, a transit depth is extracted.
 
Due to thermo-elastic dilatation of the platform during observations and the kinematic aberration, stars in the PLATO focal plane can drift horizontally by up to approximately 1.3 pixels per quarter, corresponding to the very maximum drift at the edge of the field of view \citep{Samadi_2019}. For a transit duration of 30~hours, this horizontal drift scales from 1.3~pixels over three months to about 0.014~pixels over 30~hours.

\begin{figure}
   \centering
   \includegraphics[width=\columnwidth]{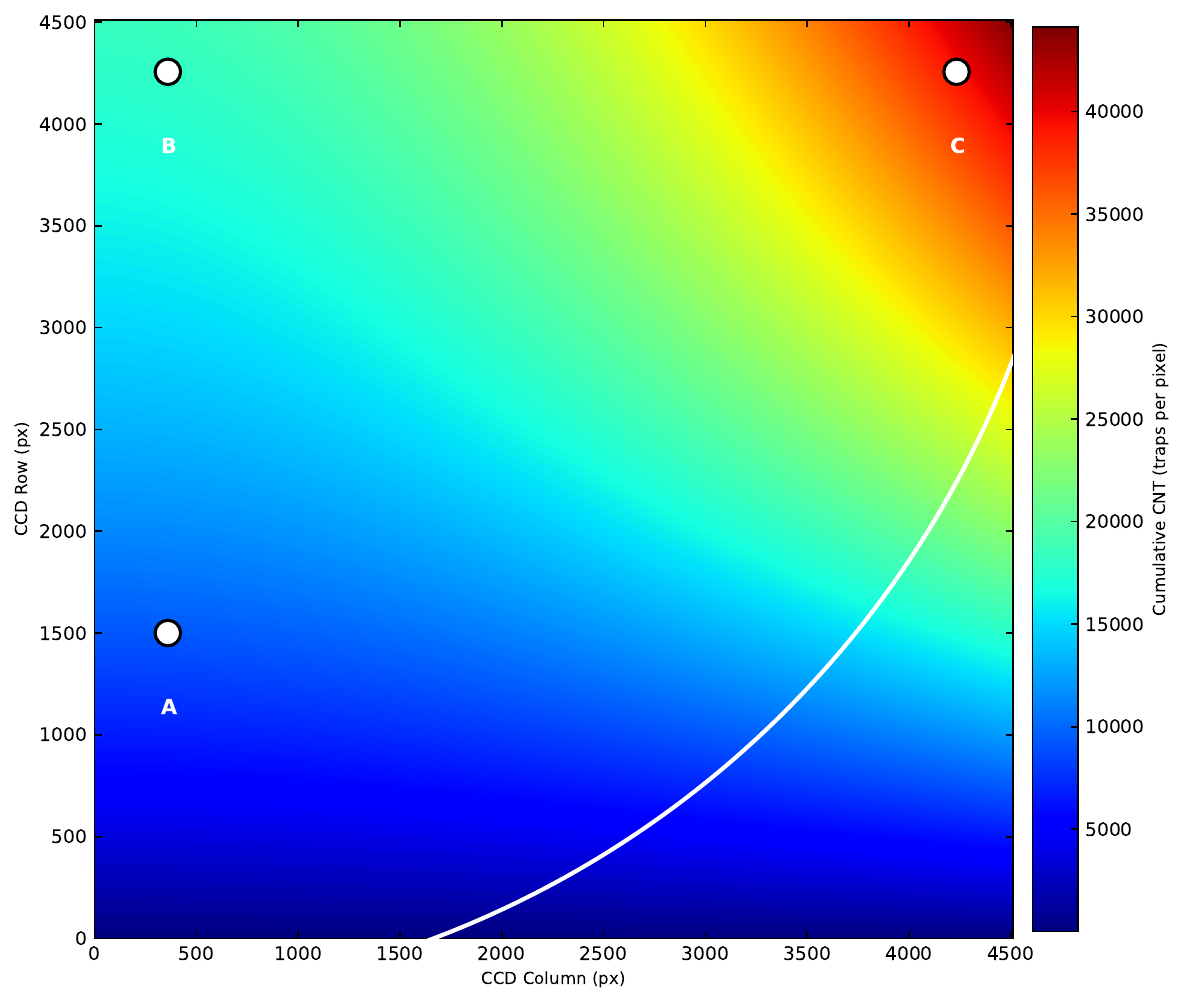}
   \caption{Map of cumulative trap density derived from Eq.~\ref{eq:cnt_true} for all species. Three positions are indicated: A (bottom left), B (top left), and C (top right). The overplotted circle with a radius of 4900 pixels represents the observable field of view.}
   \label{fig:trap_density_ccd}
 \end{figure}

We calculated the bias on transit depth due to CTI for cases corresponding to nominal mission lifetime of 4 years and extended mission lifetime of 8 years. We also incorporated positional dependence, based on the imagette's location in a region of high or low CTI. We evaluated three representative CCD positions, bottom left (A), top left (B), and top right (C), corresponding to different trap density regimes derived from Eq.\ref{eq:cnt_true} (see Fig.\ref{fig:trap_density_ccd}). The bottom-left corner (A) is the region with the lowest cumulative trap density. Conversely, the top-right corner (C) has the highest cumulative trap density. The top-left corner (B) represents an intermediate case.

To quantify the impact of CTI on transit light curves, we defined the photometric bias, \(\Delta\), as the difference between the pristine transit depth (no CTI) and the CTI-affected depth, expressed in both pixels-per-million (ppm) and percentage. Table~\ref{table:table_2} reports the impact of CTI on the transit depth of an Earth-sized planet transiting a solar-type star with $m_V = 11$ under different mission durations and trap density scenarios. For completeness, it also presents the corresponding bias for a bright star ($m_V = 9$) and a faint star ($m_V = 13$) in a region of high CTI (position C) during the extended 8-year mission. It is worth noting that these calculations were done for the worst CCD (among all the cameras), i.e. for the CCD with the highest maximum in the TNID.

\begin{table}[h!]
\centering
\caption{CTI depth across mission durations, detector positions, and magnitudes.}
\label{table:table_2}
\renewcommand{\arraystretch}{1.2}
\setlength{\tabcolsep}{3pt}
\begin{tabular}{cccccc}
\hline\hline
Years & Position & V & CTI depth (ppm) & $\Delta$ (ppm) & $\Delta$ (\%) \\
\hline
\multirow{3}{*}{8} 
   & C & 11 & 109.979 & 4.179 & 3.95 \\
   & B & 11 & 107.882 & 2.082 & 1.97 \\
   & A & 11 & 106.925 & 1.125 & 1.06 \\
\hline
\multirow{3}{*}{4} 
   & C & 11 & 108.235 & 2.435 & 2.30 \\
   & B & 11 & 106.912 & 1.112 & 1.05 \\
   & A & 11 & 106.383 & 0.583 & 0.55 \\
\hline
\multirow{2}{*}{8} 
   & C & 9  & 109.131 & 3.331 & 3.15 \\
   & C & 13 & 107.178 & 1.378 & 1.30 \\
\hline
\end{tabular}
\tablefoot{The pristine transit depth is equal to~\(105.8\)~ppm (typical value for an Earth-sized planet orbiting a solar-type star).}
\end{table}

\begin{figure}
 \centering
 \includegraphics[width=\columnwidth]{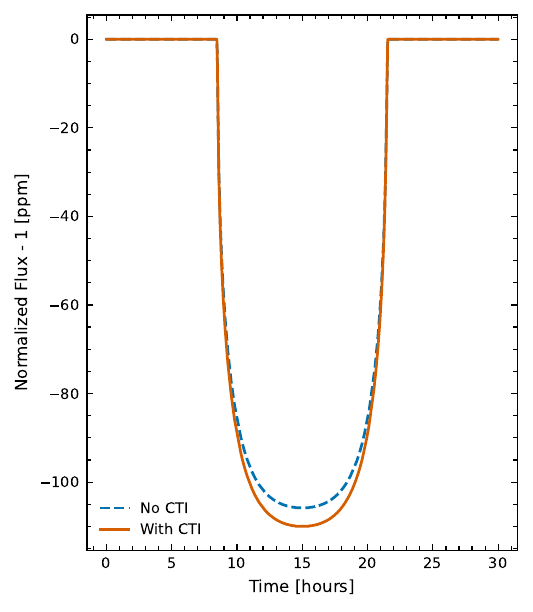}
\caption{Comparison of transit light curves for the worst-case scenario: an 8-year mission at high trap density (position~C) and \(m_V = 11\). The solid orange curve includes CTI effects, while the dashed blue curve shows the CTI-free case. CTI-induced charge loss introduces a photometric bias of \(3.95\%\).}
  \label{fig:transit_depth_bias}
\end{figure}

The pristine and CTI-affected transit depths associated with an Earth-sized planet orbiting a solar-type star are shown in Fig.~\ref{fig:transit_depth_bias} for the worst-case scenario, corresponding to an 8-year mission and an imagette located in a high-CTI region (position~C). The photometric bias is clearly visible: CTI causes an overestimation of the transit depth.  

The origin of this bias is the CTI-induced charge loss, which does not scale linearly with the stellar flux. As a result, the deferred charge release affects the in-transit and out-of-transit measurements differently, thereby influencing the measured transit depth. This non-linear behaviour is the root cause of the photometric bias seen in the observed light curve.

We found that the horizontal drift had no impact on the bias, as the light curves had been de-trended beforehand. This will also be the case for the PLATO light curves (however, see the discussion in Sect.~\ref{subsec:discussion}).

PLATO science requirements dictate that the planet-to-star radius ratio (\(R_{\mathrm{p}}/R_{\mathrm{s}}\)) must be measured with a relative precision better than \(2\%\) for an Earth-sized planet transiting a G-type star (goal: \(m_V = 11\); \citealt{PLATO2019SciRD}). Because the transit depth (\(\delta\)) is proportional to the square of this ratio, standard error propagation yields
\begin{equation}
 \frac{\Delta\!\left(R_{\mathrm{p}}/R_{\mathrm{s}}\right)}
    {R_{\mathrm{p}}/R_{\mathrm{s}}}
 \simeq \frac{1}{2}\,
     \frac{\Delta\delta}{\delta}\, .
 \label{eq:radius_ratio_transit_depth}
\end{equation}

Our worst-case CTI bias of \(3.95\%\) in transit depth corresponds to a \(1.98\%\) bias in the recovered radius ratio. This value alone consumes the entire \(2\%\) accuracy budget shared among all instrumental and astrophysical error sources, underscoring the necessity for accurate CTI correction. To test the effect of transit depth, we also computed the bias for the deeper transits of Jupiter-sized planets and found a similar value of \(3.91\%\). This similarity suggests that the CTI-induced bias does not strongly depend on the transit depth.

To evaluate the role of the fastest-releasing species, we compared this baseline bias of \(3.95\%\) with the case where the two fastest species are excluded. When the first and second species are excluded, the bias reduced to \(3.63\%\), which is a departure of \(0.32\%\) from the baseline value. This indicates that the fastest traps, particularly the first species, contribute negligibly to the systematic bias introduced by CTI.

We then investigated the impact of the third species by excluding it along with the first and second species. In this case, the bias decreased further to \(2.21\%\), a significant change of \(1.74\%\) from the baseline bias, suggesting that the third species plays a major role in the CTI-induced bias, alongside the slowest fourth species.

\section{CTI calibration strategy}
\label{sec:cti_calibration_strategy}
\begin{figure}
  \centering
  \includegraphics[width=\columnwidth]{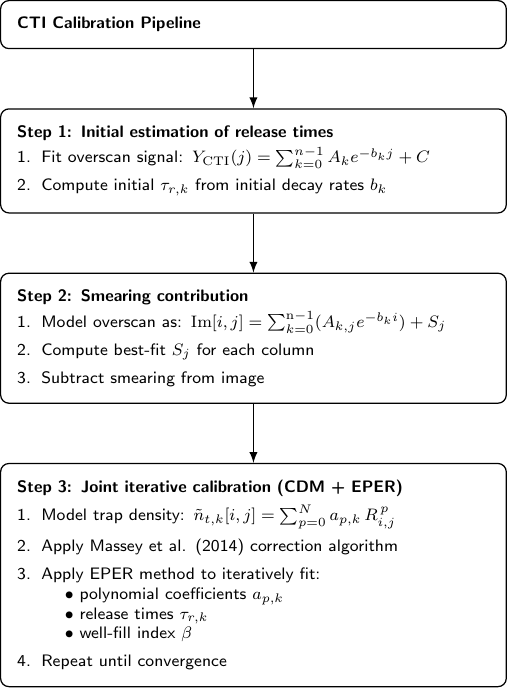}
\caption{Workflow of the CTI calibration strategy: initial estimates for the release time constants (\(\tau_{\mathrm{r,k}}\)) are obtained by fitting an empirical model to the overscan signal; smearing is corrected via a column-wise exponential-plus-constant fit; the mean trap density is modelled as a quadratic polynomial to reproduce its radial dependence; and the EPER method is applied, in conjunction with a CTI correction algorithm, to iteratively fit the polynomial coefficients and the well-fill power-law index (\(\beta\)), and to refine the release time constants (\(\tau_{\mathrm{r,k}}\)).}
\label{fig:calibration_strategy_summary}
\end{figure}

This section outlines the calibration strategy employed in this study, which relies on full or partial images acquired at the beginning of each quarter (i.e. every three months). First, we defined an empirical model to determine the number of trap species and to derive initial estimates for their release time constants (\(\tau_{\mathrm{r,k}}\)). Next, we addressed and removed the contribution from smearing. Finally, we adopted a polynomial model to describe the spatial variation in the trap density, reproducing the radial dependence discussed in Sect.~\ref{subsec:radiation_analysis}. We then applied the EPER method \citep{Prodhomme_2016} to constrain the polynomial coefficients of the trap density model (\(a_{\mathrm{p,k}}\)), along with the \(\beta\) parameter and the release time constants (\(\tau_{\mathrm{r,k}}\)). The overall procedure is summarised in Fig.~\ref{fig:calibration_strategy_summary}.

\subsection{Calibration dataset}
\label{subsec:calibration_dataset}
As PLATO is not yet in-flight, we used a simulated CCD image containing a stellar field for the calibration procedure. This image is generated with PLATOSim \citep{Jannsen_2024}, a simulation tool designed to produce photometric time series and CCD images tailored to the PLATO mission. The simulated image includes several sources of noise---smearing, photon noise, and readout noise---to approximate realistic observational conditions. To improve the signal-to-noise ratio (S/N), we averaged over 10 exposures. The upper limit of 10 exposures is decided by the in-flight operational constraints (specifically to minimise the duration of the calibration phase and the associated telemetry volume).

We simulated forward CTI effects using the cumulative trap distribution described in Sect.~\ref{subsec:radiation_analysis}. The distribution follows an axisymmetric radial profile, scaled to match the radiation dose for PLATO's extended 8-year mission scenario \citep{ohb2024radiation}. The mean trap densities are scaled accordingly to construct the cumulative map, centred on the optical axis of the CCD.

\subsection{Overscan analysis and CTI signature}
\label{subsec:overscan_analysis_cti_signature}
\begin{figure}
  \centering
  \includegraphics[width=\columnwidth]{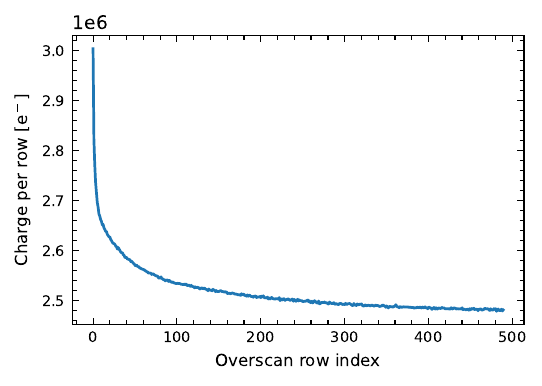}
\caption{Sum of the signal in each overscan row across all columns as a function of row index. The profile follows an exponential decay, with physical rows located before overscan row index~0.}
  \label{fig:overscan_row_signal}
\end{figure}

Parallel overscan rows are virtual rows beyond the CCD's physical imaging area, accessed after the last physical row is read out. Hereafter, `overscan' refers to this parallel overscan region. These rows contain deferred charges that were not recorded during readout because of trap capture, providing a diagnostic window to constrain the CTI signature. Figure~\ref{fig:overscan_row_signal} shows the overscan signal, summed across all the columns for a given row index. Its shape reflects the delayed release of charge by different trap species: those with short release times dominate near the start, whereas slower species extend further into the overscan region. The slowest traps are of primary concern, as their release occurs beyond the photometric aperture. 

\subsection{Initial estimates for the trap release time (Step 1)}
\label{subsec:trap_release_time}
From the profile of the signal in Fig.~\ref{fig:overscan_row_signal}, a multi-exponential function is used to  (empirically) model the signal contained in the overscan region. It is illustrated in the Eq.~\ref{eq:rt_model_equation} as

\begin{equation}
  Y_{\mathrm{CTI}}(i) = \sum_{k=0}^{n-1} A_k e^{-b_k i} + C\, , 
\label{eq:rt_model_equation}
\end{equation}

\noindent where \( A_k \) is the amplitude for trap species \( k \), corresponding to the range (max - min) of the signal; \( b_k \) is the decay rate for species \( k \), inversely related to the release time (see Eq.~\ref{eq:rt_decay_rate_relationship}); \( C \) is the constant offset (sum of smearing in each individual column); \( i \) is the row index along the CCD column (parallel transfer direction); and \( n \) is the total number of trap species; \( Y_{\text{CTI}}(i) \) is the signal level (in e\textsuperscript{-}/pix) at row \( i \), including CTI trailing effects. The relationship between the decay rate (\( b_k \)) and release time (\( {\tau}_{r,k}\)) is given by

\begin{equation}
  {\tau}_{r,k} = \frac{1}{b_k} \times t_{\text{transfer\ cadence}}\, .
 \label{eq:rt_decay_rate_relationship}
\end{equation}

\begin{figure}
\centering
\includegraphics[width=\columnwidth]{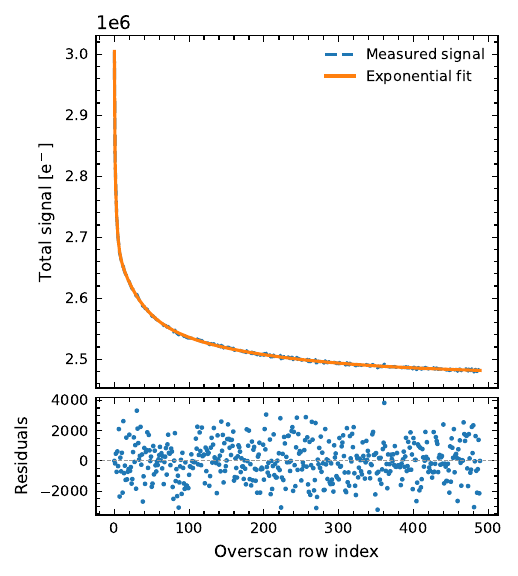}
\caption{First estimates of the release time constants obtained by fitting the overscan signal. \textit{Top}: Measured signal plotted against the exponential model (Eq.~\ref{eq:rt_model_equation}) as a function of row index. \textit{Bottom}: Residuals, indicating good agreement between the data and the model.}

\label{fig:release_time_results}
\end{figure}

We designed an iterative fitting strategy that progressively increases the number of species until the solution converges within a predefined threshold. Applying this strategy to our simulated image yielded a maximum of four trap species, with a final reduced \(\chi^{2}\) of \(1.141\).
The first estimates for the release time values and decay constants for these species, ordered from fastest to slowest, are listed in Table~\ref{table:table_4}. We provide the associated uncertainties for the release times and classify each species according to its characteristic release timescale: fast, intermediate, or slow. The corresponding fit results, along with the residuals, are presented in Fig.~\ref{fig:release_time_results}. The difference between our best-fit release time values (Table~\ref{table:table_4}) and the reference values from \citet[see our Table~\ref{table:table_1}]{Prodhomme_2016} arises from the distinct methodologies employed in each case. Our simulations include a realistic stellar field, which influences the empirical model used to derive best-fit release times, unlike the flat-field image adopted by \citet{Prodhomme_2016}. In the forthcoming Sect.~\ref{subsec:trap_density}, we iteratively refit the trap release times, using these values (Table~\ref{table:table_4}) as initial guesses.

\begin{table}
\caption{First estimates of decay and release time constants obtained during Step~1.}
\label{table:table_3}
\centering
\begin{tabular}{ccccc}
\hline\hline
Species & Type & \(b_k\) & \(\tau_{\mathrm{r},k}\) & Uncertainty \\
        &      & (pix\(^{-1}\)) & (s) & (s) \\
\hline
1 & Fast          & \(2.87 \times 10^{1}\) & \(3.10 \times 10^{-5}\)  & \(-\) \\
2 & Interm. & \(4.39 \times 10^{-1}\) & \(2.03 \times 10^{-3}\)  & \(\pm 4.26 \times 10^{-5}\) \\
3 & Interm. & \(3.37 \times 10^{-2}\) & \(2.64 \times 10^{-2}\)  & \(\pm 5.44 \times 10^{-4}\) \\
4 & Slow          & \(5.68 \times 10^{-3}\) & \(1.57 \times 10^{-1}\)  & \(\pm 3.44 \times 10^{-3}\) \\
\hline
\end{tabular}
\tablefoot{The species are ordered from fastest to slowest. Uncertainties on the release time constants are given at the \(1\sigma\) level. The qualitative classification is based on the characteristic release timescale. The decay constant (\(b_k\)) is obtained from an exponential fit to the overscan rows and is related to the species-specific release time (\(\tau_{\mathrm{r},k}\)) via Eq.~\ref{eq:rt_decay_rate_relationship}. Uncertainty for Species~1 is marked with a dash as it cannot be constrained because its release time is shorter than the line transfer time.}
\end{table}

\subsection{Smearing estimation and removal (Step 2)}
\label{subsec:smearing}
\begin{figure}
  \centering
  \includegraphics[width =\columnwidth]{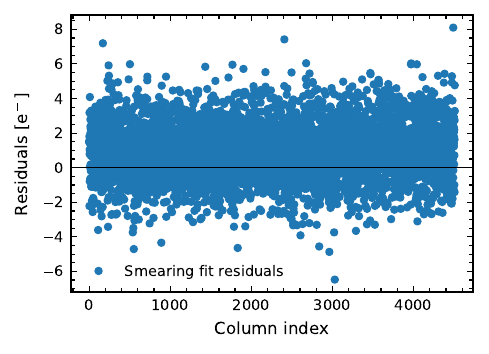}
\caption{Residuals from the smearing fit described in Sect.~\ref{subsec:smearing}, shown as a function of column index.}
  \label{fig:smearing_residuals}
\end{figure}

Smearing occurs when pixels accumulate additional charge during line transfer, leading to streaks extending from sources along the CCD column (in both directions). This step estimates the smearing contribution in each overscan column and removes it prior to further processing. The signal in each overscan column is modelled as
\begin{equation}
\mathrm{Im}[i,j] = \sum_{k=0}^{n-1} \left(A_{k,j} e^{-b_k i}\right) + S_j\, ,
\label{eq:smearing_fit_model}
\end{equation}
where \(S_{j}\) represents the constant smearing term for column \(j\) and \(A_{k, j}\) is a free parameter. The decay constants \(b_k\) are taken from Sect.~\ref{subsec:trap_release_time}. We determined \(A_{k,j}\) and \(S_j\) via a least-squares fit and subtracted \(S_j\) from the image (including from the parallel overscan rows) to remove the smearing. We display the residuals obtained from this step in Fig.~\ref{fig:smearing_residuals}. We obtained a dispersion value of \(1.6\) units from the fit.

\subsection{Trap density calibration and radial dependence (Step 3)}
\label{subsec:trap_density}
The iterative strategy described in Sect.~\ref{subsec:trap_release_time} determines the number of species to consider and their corresponding release times. The next stage of calibration involved estimating the cumulative trap density via polynomial coefficients, after smearing removal (see the previous section).

This stage used the EPER method \citep{Prodhomme_2016} for PLATO to analyse the overscan region. EPER relies on the CDM: the deferred charge appearing as trails in the overscan is modelled and fitted to the signal in the overscan region in order to infer the CTI parameters. From this, we derived the cumulative trap density for each species, \(\mathrm{cnt}_{t,k}\), and constructed a model based on the best-fit values.

Based on the axisymmetric radial dependence inferred from the work of \citet{ohb2024radiation} and covered in Sect.~\ref{subsec:radiation_analysis}, we modelled the mean trap density as a quadratic polynomial function:
\begin{equation}
  \tilde{n}_{t,k}[i, j] = \sum_{p=0}^{N} a_{p,k}\, R_{i,j}^{\,p}\, , 
  \label{eq:mean_trap_density_model_eq}
\end{equation}
where \(p\) denotes the degree of the polynomial, \(R\) denotes the normalised radial distance from the optical centre (see Eq.~\ref{eq:Rnorm}), and \(i\) and \(j\) represent the row and column indices of the CCD, respectively. From Eq.~\ref{eq:mean_trap_density_model_eq}, we can similarly model the cumulative trap density as

\begin{equation}
\begin{aligned}
  \tilde{\mathrm{cnt}}_{k}[i, j] &= \sum_{i'=1}^{i} \tilde{n}_{t,k}[i', j] \\
  \tilde{\mathrm{cnt}}_{k}[i, j] &= \sum_{p=0}^{N} a_{p,k}\,\left(\sum_{i'=0}^{i}R_{i', j}^{\,p}\right)\, .
\end{aligned}
\label{eq:cnt_model_eq}
\end{equation}

Equation~\ref{eq:cnt_model_eq} expresses the cumulative number of traps (\(\tilde{\mathrm{cnt}}_{k}[i, j]\)) for species \(k\) at any given pixel \((i, j)\) as the sum of the local trap densities (\(\tilde{n}_{t,k}[i', j]\)) integrated along the parallel transfer direction up to row \(i\). The specific case of \(i = 4510\) corresponds to the total number of CCD rows, representing full-column integration across the detector height. The cumulative number of traps is directly related to the radial dependence expressed by the polynomial coefficients (\(a_{p,k}\)) from Eq.~\ref{eq:mean_trap_density_model_eq}.

We implemented an iterative calibration algorithm applied simultaneously to the CCD image. The impact of CTI, and consequently the charge trails in the overscan rows, is primarily caused by the capture of charges in pixels located within a limited distance from the top of the CCD. Therefore, it is not necessary to model all the CCD rows, as doing so would significantly increase the computational cost without improving accuracy. To optimise the calibration process, we determined the number of science rows that must be considered by analysing the characteristic length (\(\mathrm{CL}\)) of the effect of the slowest trap species. This characteristic length \(\mathrm{CL}\) (in pixels) is defined as
\begin{equation}
    \begin{aligned}
      \mathrm{CL} = \frac{1}{b_{k}^{\mathrm{slowest}}}\, ,
    \end{aligned}
    \label{eq:cl_definition}
\end{equation}
where \(b_{k}^{\mathrm{slowest}}\) refers to the decay rate associated with the slowest trap species computed in Step 1 (Sect.~\ref{subsec:trap_release_time}). We obtained \(CL \simeq  180\) pixels from the value obtained for \(b_{k}^{\mathrm{slowest}}\) (Table~\ref{table:table_3}).
Traps located at distances significantly greater than the characteristic length \(CL\) from the top of the CCD do not contribute to the deferred charge observed in the overscan region. Therefore, it is sufficient to restrict the computation to a vertical distance of approximately three times this length. In this work, we defined the effective characteristic length as \(\mathrm{CL}_{\mathrm{eff}} = 3 \times \mathrm{CL}\).

For the CCD image part spanning the sum of \(\mathrm{CL}_{\mathrm{eff}}\) and the overscan region, the \citet{Massey_2014} correction is first applied to estimate the pixel intensities in the absence of CTI. This correction is used solely to initialise the estimation of the polynomial coefficients used to model the cumulative trap density, \(\mathrm{cnt}_{k}\), as part of the calibration process. The procedure uses the \texttt{run\_cdm\_cumul\_radia\_poly} routine (see the Python implementation in \href{https://github.com/mishrashaunak/plato-cti-correction}{plato-cti-correction}), which requires the current estimates of the polynomial coefficients (\(a_{p,k}\)) for the working region, along with all the other constant CTI parameters.

Using the corrected pixel values over the full working region (i.e. \(\mathrm{CL}_{\mathrm{eff}} +\) overscan), we forward-modelled CTI trailing to generate a synthetic `model' image, including the overscan. A cost function is then defined over the overscan region to optimise the polynomial coefficient map (\(a_{p,k}(i,j)\)), the release time constants (\(\tau_{\mathrm{ r,k}}\)), and the well-fill power-law index (\(\beta\)), by minimising the difference between the measured deferred charge and that predicted by the CDM (EPER method).

The release time associated with the first species has been fixed to \(\tau_{\mathrm{r,0}} = 3.10 \times 10^{-5}\,\mathrm{s}\), corresponding to the value obtained in Step~1 (Table~\ref{table:table_3}). This choice is motivated by the fact that \(\tau_{\mathrm{r,0}}\) is shorter than the line transfer time (\(\sim 900~\mu\mathrm{s}\)), and therefore cannot be constrained by the fit.

The cost function and the CTI correction algorithm were combined in an iterative process: at each step, the corrected working region is updated, and the least-squares fit is repeated until the improvement in the reduced \(\chi^{2}\) falls below \(0.001\). To accelerate computations, we parallelised this process and selected every $N_c$ column within the working region to reduce the computational load.

To speed up convergence, we first ran our step 3 algorithm using only \(1/3\) of \(\mathrm{CL}_{\mathrm{eff}}\) and set \(N_c = 10\). Then, in a second and final iteration, we used all \(\mathrm{CL}_{\mathrm{eff}}\) rows and adopt \(N_c = 5\). This second iteration is treated as the last one since we observed no improvement beyond the reduced \(\chi^2\) threshold. We noted that increasing \(\mathrm{CL}_{\mathrm{eff}}\) further or decreasing \(N_c\) did not significantly change the results.

There is a total of 12 polynomial coefficients: three (for a quadratic function) per species across four trap species. The fit was initialised using the best-fit release times from Table~\ref{table:table_3}, and a uniform starting guess of \(1.0\) was adopted for each coefficient of the radial polynomial associated with the four trap species. We specified an upper bound of \(+\infty\) for all free parameters (except \(\beta\)). The lower bounds for coefficients \(a_0\) and \(a_2\) were set to zero, as these terms are not expected to be negative. However, the lower bound for the linear term \(a_1\) was set to \(-\infty\), motivated by the presence of a local maximum at the centre of the focal plane in Fig.~\ref{fig:camera_9_radiation_map}, followed by a decrease and then a rise towards the edges. This pattern justified allowing \(a_1 < 0\).

\begin{figure}
  \centering
  \includegraphics[width=\columnwidth]{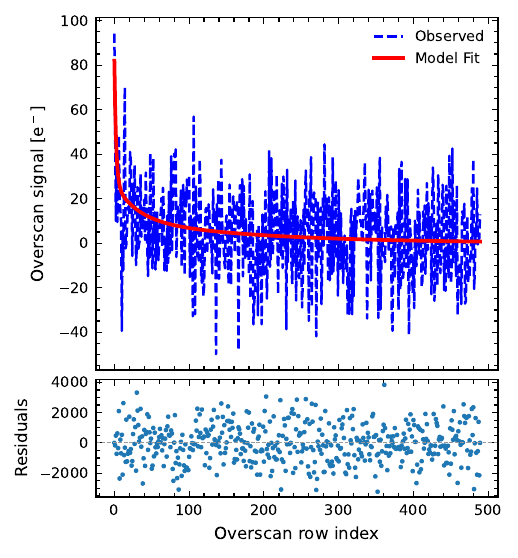}
\caption{\textit{Top}: Iterative EPER-based fit for \(\beta\), release time constants (\(\tau_{\mathrm{r, k}}\)), and polynomial coefficients (\(a_{\mathrm{p,k}}\)) representing the trap densities in the overscan region for column index \(2255\). \textit{Bottom}: Residuals.}
\label{fig:trap_density_fit_rep_column}
\end{figure}

For the well-fill power index (\(\beta\)), we used an initial guess of 0.5---representing the midpoint of its expected range between 0 and 1---and constrained it within the optimiser using bounds of 0 (lower) and 1 (upper). The initial guess for each release time (\(\tau_{\mathrm{r},k}\)) corresponded to the values derived during Step~1 (see Table~\ref{table:table_3}), with bounds of \(0\) and \(+\infty\) applied for the fitting. 
This iterative scheme yielded a map of the species-specific polynomial coefficients (\(a_{p,k}\)), as well as fitted values for the release times (\(\tau_{\mathrm{r},k}\)) and \(\beta\), along with associated \(1\sigma\) uncertainty maps. The final fit achieved a reduced \(\chi^{2}\) of \(1.223\). Figure~\ref{fig:trap_density_fit_rep_column} shows an example fit for a representative column (index 2255), located near the centre of the CCD.

We report the fitted values of the polynomial coefficients (\(a_{p,k}\)), the release time constants (\(\tau_{\mathrm{r},k}\)), and \(\beta\) in Table~\ref{table:table_4}, along with their respective \(1\sigma\) uncertainties. The fitted value of \(\beta\) is found to be very close to the true value from \citet{Prodhomme_2016}. The release time constants (\(\tau_{\mathrm{r},k}\)) for species 3 and 4 are also estimated more accurately here than in Step~1, as shown in Table~\ref{table:table_4}.

\subsection{Polynomial coefficients from theory and fit}
\label{subsec:polynomial_coefficients_theory_fit}

The fitted polynomial coefficients in Eq.~\ref{eq:mean_trap_density_model_eq} can be directly related to their theoretical counterparts. From Eqs.~\ref{eq:radial_relationship} and \ref{eq:nt_true}, these coefficients are expressed as
\begin{equation}
  \begin{aligned}
    a_{0,\mathrm{k}}^{\mathrm{theory}} &= n_{t,k}^{\mathrm{scaled}} \times \left(\frac{D_{\mathrm{min}}}{D_{\mathrm{max}}}\right) \\
    a_{1,\mathrm{k}}^{\mathrm{theory}} &= 0 \\
    a_{2,\mathrm{k}}^{\mathrm{theory}} &= n_{t,k}^{\mathrm{scaled}} \times \left(1 - \frac{D_{\mathrm{min}}}{D_{\mathrm{max}}}\right)\, , 
  \end{aligned}
  \label{eq:theoretical_polynomial_coefficients}
\end{equation}
where \(a_0\) represents the constant term and \(a_2\) the quadratic term. The linear term \(a_1\) is zero by definition. We provide the theoretical values for the \(a_0\), \(a_1\), and \(a_2\) coefficients in Table~\ref{table:table_4} alongside their fitted counterparts.

\begin{table}
\renewcommand{\arraystretch}{1.2}
\setlength{\tabcolsep}{4pt}
\caption{Fitted and theoretical polynomial coefficients ($a_0, a_1$, and $a_2$) per species.}
\centering
\begin{tabular}{ccccc}
\hline\hline
Species & Parameter & Fit & Theory & Uncertainty \\
\hline
\multirow{3}{*}{1} 
  & \(a_0\) & \(4.23 \times 10^{-3}\) & 0.754 & \(6.27 \times 10^{5}\) \\
  & \(a_1\) & \(-15.0\) & 0.000 & \(1.19 \times 10^{7}\) \\
  & \(a_2\) & \(2.56\) & 3.882 & \(1.92 \times 10^{6}\) \\
\hline
\multirow{4}{*}{2} 
  & \(a_0\) & \(0.166\) & 0.120 & \(4.03 \times 10^{-2}\) \\
  & \(a_1\) & \(0.155\) & 0.000 & \(1.72 \times 10^{-1}\) \\
  & \(a_2\) & \(0.743\) & 0.618 & \(1.63 \times 10^{-1}\) \\
  & \(\tau_{\mathrm{r},2}\) & \(1.34 \times 10^{-3}\) & \(2.0 \times 10^{-3}\) & \(1.47 \times 10^{-5}\) \\
\hline
\multirow{4}{*}{3} 
  & \(a_0\) & \(0.274\) & 0.255 & \(4.97 \times 10^{-2}\) \\
  & \(a_1\) & \(-0.164\) & 0.000 & \(2.07 \times 10^{-1}\) \\
  & \(a_2\) & \(1.59\) & 1.311 & \(1.91 \times 10^{-1}\) \\
  & \(\tau_{\mathrm{r},3}\) & \(2.27 \times 10^{-2}\) & \(2.4 \times 10^{-2}\) & \(2.19 \times 10^{-4}\) \\
\hline
\multirow{4}{*}{4} 
  & \(a_0\) & \(1.12\) & 1.018 & \(7.88 \times 10^{-2}\) \\
  & \(a_1\) & \(-0.243\) & 0.000 & \(3.27 \times 10^{-1}\) \\
  & \(a_2\) & \(5.65\) & 5.244 & \(3.01 \times 10^{-1}\) \\
  & \(\tau_{\mathrm{r},4}\) & \(1.46 \times 10^{-1}\) & \(1.4 \times 10^{-1}\) & \(5.15 \times 10^{-4}\) \\
\hline
\multirow{1}{*}{--} 
  & \(\beta\) & \(0.376\) & \(0.370\) & \(5.67 \times 10^{-4}\) \\
\hline
\end{tabular}
\label{table:table_4}
\tablefoot{This table includes the true values of $\beta$ and $\tau_{\mathrm{r},k}$ from \citet[see our Table~\ref{table:table_1}]{Prodhomme_2016} and reports the fitted values together with their respective $1\sigma$ uncertainties. The release time for the first species was fixed to $\tau_{\mathrm{r},0} = 3.10 \times 10^{-5}\,\mathrm{s}$ (from Step~1; Table~\ref{table:table_3}), as it is shorter than the line transfer time ($\sim 900~\mu\mathrm{s}$) and cannot be constrained by the fit.}
\end{table}

From the fitted polynomial coefficients obtained in Step~3 (Sect.~\ref{subsec:trap_density}), we derived the cumulative trap density at the top of the CCD, i.e. the quantity \(\tilde{\mathrm{cnt}}_{k}[4509, j]\) for columns indexed from \(j = 0\) to \(j = 4059\). This estimate can then be directly compared to the simulated (or true reference) cumulative trap density defined by Eq.~\ref{eq:cnt_true}.

\begin{figure*}
\centering
\includegraphics[width=\textwidth]{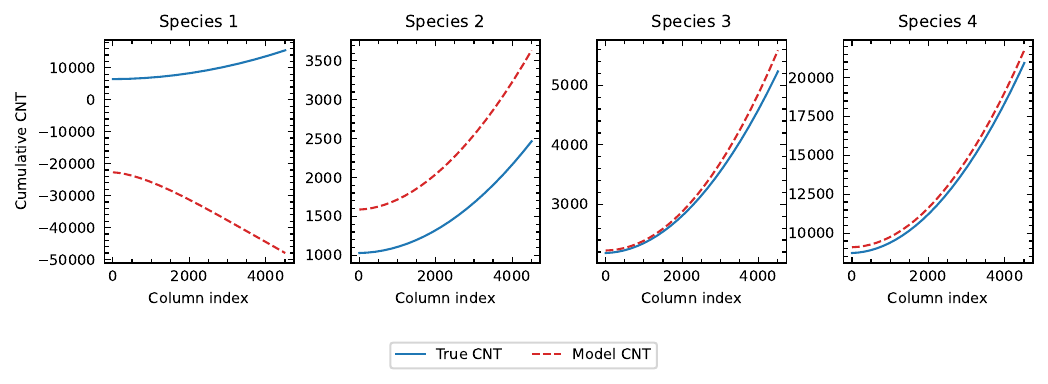}
\caption{Comparison of cumulative trap density (\(\mathrm{cnt}\)) at the top of the CCD for all species from two sources: the simulation ground truth (in blue) and the fitted polynomial model (in dashed red). The plot highlights certain species-specific behaviour, with Species~2 exhibiting a systematic offset. Note that Species~1 is poorly constrained because its release time is significantly shorter than the line transfer time; however, as detailed in Sect.~\ref{subsec:bias_transit_depth_measurements}, this species contributes negligibly to the total CTI-induced bias.}
\label{fig:step3_results}
\end{figure*}

Figure~\ref{fig:step3_results} illustrates these two cases overplotted for all four species. The result for the first species is poorly constrained, while the second species is mildly constrained and shows a slight systematic offset relative to the true profile. However, as noted previously in Sect.~\ref{subsec:bias_transit_depth_measurements}, the first two species have negligible impact on the CTI-induced bias. Accordingly, the discrepancies found for the first and second species have no consequences. Ultimately, these calibrated coefficients, capturing the spatial dependence of the trap density, form the basis for the CTI correction procedure described in the next section.

\section{CTI correction strategy}
\label{sec:cti_correction_strategy}
In this section we describe a strategy to correct CTI in imagettes acquired with the normal cameras at a 25\,s cadence. The correction builds on the algorithm of \citet{Massey_2014} and uses our calibrated trap-density map.

\subsection{Correction procedure}
\label{subsec:correction_procedure}

The fitted polynomial coefficients obtained in the previous section provided the basis for constructing the trap density map necessary for CTI correction of the planetary transit discussed in Sect.~\ref{subsec:bias_transit_depth_measurements}. Using Eq.~\ref{eq:theoretical_polynomial_coefficients}, we defined the per-pixel trap densities across the full CCD area as
\begin{equation}
  n_{t,k}[i, j]^{\mathrm{full}} = a_{0,k}^{\mathrm{fit}} + a_{1,k}^{\mathrm{fit}} \times R_{i,j} + a_{2,k}^{\mathrm{fit}} \times R_{i,j}^2\, ,
  \label{eq:model_polynomial_coefficients}
\end{equation}
where \(a_{0,k}^{\mathrm{fit}}\), \(a_{1,k}^{\mathrm{fit}}\), and \(a_{2,k}^{\mathrm{fit}}\) are the fitted polynomial coefficients for species \(k\) (see Table~\ref{table:table_4}), and \(R_{i,j}\) is the normalised radial coordinate at pixel \((i,j)\)
as defined in Eq.~\ref{eq:Rnorm}. The cumulative number of traps per pixel is computed by summing \(n_{t,k}[i,j]\) along the CCD rows (see Eq.~\ref{eq:cnt_model_eq}). This cumulative trap density map forms the basis for correcting CTI-affected planetary transits within the imagette containing the source.

Imagettes acquired during observations are small (6\(\times\)6 pixels), but charges deferred by CTI from sources read out before the imagettes can enter its area, introducing contamination. In order to take this into account, it is necessary to know the filling history of the traps, which is determined by the charge distribution in pixels read out before the imagettes along the readout direction. To account for this, we used the full-frame images acquired at the start of each quarter, which provide the flux distribution outside the imagettes.

The CTI correction is implemented using an adaptation of the iterative scheme proposed by \citet{Massey_2014}, applied to each imagette. The algorithm employed the routine \texttt{run\_cdm\_parallel\_cumul}, using the cumulative number of traps (\(\mathrm{cnt}_k[i,j]\)) as input (together with all the other constant CTI parameters). Each column is iteratively corrected for a maximum of 10 iterations or until the reduced \(\chi^{2}\) fell below a threshold of \(0.001\). After correction, photometry was extracted from the imagettes using an optimal aperture mask, following the procedure described in Sect.~\ref{subsec:bias_transit_depth_measurements}.

\subsection{Correction results  and performance evaluation}
\label{subsec:correction_results_performance_evaluation}
Following the procedure detailed in preceding section and using the fitted polynomial coefficients, we now demonstrate the results of our correction strategy.

The final transit depths (in ppm) were computed for each case. We defined the residual correction as
\begin{equation}
\mathrm{Residual} = \big| D_{\mathrm{no\ CTI}} - D_{\mathrm{corr}} \big|\, ,
\label{eq:residual_corr_TD}
\end{equation}
where \(D_{\mathrm{no\ CTI}}\) denotes the transit depth without CTI, and \(D_{\mathrm{corr}}\) denotes the transit depth after CTI correction. This framework enabled us to assess the performance of the CTI correction strategy in recovering the true transit depth and, more broadly, the overall efficiency of the method. We summarise the CTI-induced bias, the residuals after correction, and the corresponding improvement in Table~\ref{table:table_5}. Results are presented for a star with \(m_V = 11\) at positions A, B, and C (see Fig.~\ref{fig:trap_density_ccd}) for mission durations of 8 and 4 years. For completeness, we also include two extreme cases: a bright star (\(m_V = 9\)) and a faint star (\(m_V = 13\)), both at position~C under the 8-year scenario, which represents the worst-case CTI impact.

\begin{table}
\centering
\caption{CTI-induced bias and correction performance.}
\label{table:table_5}
\renewcommand{\arraystretch}{1.2}
\setlength{\tabcolsep}{4pt}
\begin{tabular}{cccccccc}
\hline\hline
Years & Pos & V & \multicolumn{2}{c}{Bias} & \multicolumn{2}{c}{Residual} & Improvement \\
\cline{4-7}
 & &  & (ppm) & (\%) & (ppm) & (\%) & (\%) \\
\hline
\multirow{3}{*}{8} 
   & C & 11 & 4.179 & 3.95 & 0.058 & 0.06 & 98.61 \\
   & B & 11 & 2.082 & 1.97 & 0.016 & 0.01 & 99.25 \\
   & A & 11 & 1.125 & 1.06 & 0.010 & 0.01 & 99.12 \\
\hline
\multirow{3}{*}{4} 
   & C & 11 & 2.435 & 2.30 & 0.040 & 0.04 & 98.34 \\
   & B & 11 & 1.112 & 1.05 & 0.005 & 0.01 & 99.52 \\
   & A & 11 & 0.583 & 0.55 & 0.002 & 0.00 & 99.74 \\
\hline
\multirow{2}{*}{8} 
& C & 9  & 3.331 & 3.15 & 0.063 & 0.06 & 98.11 \\
& C & 13 & 1.378 & 1.30 & 0.017 & 0.02 & 98.79 \\
\hline
\end{tabular}
\tablefoot{Results are shown for different mission durations, positions, and magnitudes. Bias represents the CTI impact, Residual shows the remaining error after correction, and Improvement quantifies the fraction of bias removed.}
\end{table}

\begin{figure}
  \centering
  \includegraphics[width=\columnwidth]{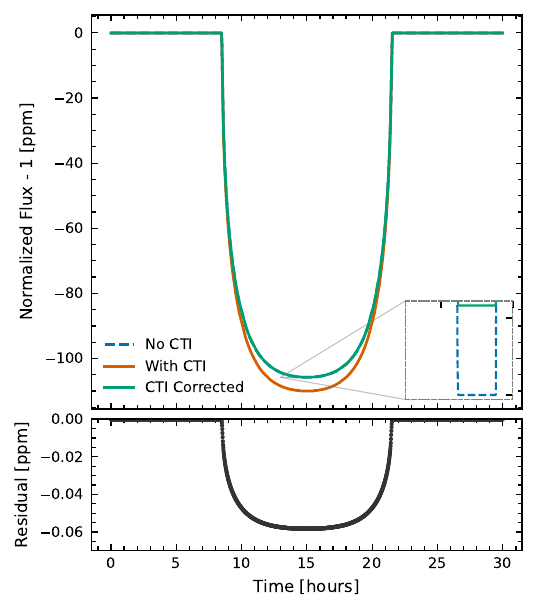}
\caption{\textit{Top}: Comparison of transit light curves for the worst-case scenario: an 8-year mission, high trap density (position~C), and \(m_V = 11\). The solid orange curve corresponds to the CTI-affected case, the solid green curve to the CTI-corrected case (nearly overlapping the dashed blue curve representing the CTI-free case). For an initial bias of \(3.95\%\), the correction reduces the residual to \(0.06\%\), achieving an improvement of \(98.61\%\). \textit{Bottom}: Correction residual. }
\label{fig:transit_depth_corrected}
\end{figure}

Figure~\ref{fig:transit_depth_corrected} compares the pristine, CTI-affected, and CTI-corrected transit depths for a star with \(m_V = 11\), representing the worst-case scenario (8-year mission duration, position C). The initial CTI bias of \(3.95\%\) was reduced to a residual of \(0.06\%\), corresponding to an improvement of \(98.61\%\), well within PLATO's compliance window.

We also quantified the error propagation introduced in the CTI correction by uncertainties in the fitted parameters using Monte Carlo simulations. Starting from the set of fitted CTI parameters, we generated random samples assuming a Gaussian distribution with zero mean and standard deviations equal to the corresponding \(1\sigma\) uncertainties. For each realisation, we computed the in-transit and out-of-transit fluxes to determine the transit depth in both the CTI-free and CTI-corrected cases. This procedure was repeated 1000 times to obtain a distribution of residuals, from which we derived a dispersion of \(0.27\%\). Increasing the number of exposures to 30 reduced the dispersion to \(0.16\%\), while including all CCD columns further lowered it to \(0.12\%\).

\section{Conclusions and discussion}
\label{sec:conclusions_discussion}
\subsection{Conclusions}
\label{subsec:conclusion}
Our aim was to assess whether PLATO can meet its scientific requirements for detecting Earth-sized planets orbiting solar-type stars via the transit method in the presence of CTI. CTI, caused by radiation-induced charge traps in the CCD substrate, leads to flux loss and redistribution, introducing systematic errors in photometric measurements. Using the radial dependence of trap density derived from the PLATO radiation analysis \citep{ohb2024radiation}, we constructed a radiation map and simulated forward CTI to quantify its impact on transit depth measurements. In the worst-case scenario, we find a systematic bias of up to \(3.95\%\), which exceeds the mission requirement of \(0.4\%\) (see Sect. \ref{sec:introduction}), underscoring the need for effective mitigation.

In response to this challenge, we designed a calibration and correction strategy for CTI. A dedicated calibration dataset was generated using PLATOSim \citep{Jannsen_2024}, simulating a realistic stellar distribution over the full CCD. To capture spatial variations in the trap density, we modelled its axisymmetric radial dependence using a polynomial function of the distance from the optical centre.
The calibration began with an empirical fit to determine the number of trap species and their initial release time estimates (\(\tau_{\mathrm{r},k}\)). We then modelled and removed the smearing contribution, which would otherwise bias subsequent steps. Next, an iterative algorithm adapted from \citet{Massey_2014}, combined with the EPER method \citep{Prodhomme_2016}, was applied to constrain, for each species, the polynomial coefficients (\(a_{p,k}\)) describing the spatial variation of trap density, as well as the well-fill power-law index (\(\beta\)) and the release time constants (\(\tau_{\mathrm{r},k}\)).

Finally, the fitted coefficients were used in a column-wise iterative correction procedure, also adapted from \citet{Massey_2014}, to approximate CTI-free images. This correction reduces the systematic bias in transit depth measurements to a level fully compliant with the mission's requirements.

\subsection{Discussion and perspectives}
\label{subsec:discussion}
Our calibration currently assumes access to the full image, including regions read out before the imagette containing the target. In practice, this assumption only holds at the beginning of a quarter, since full-frame images will not be available during the observations. Consequently, sources located in regions read out before the imagettes will drift over time, complicating CTI correction.

Another limitation arises from our reliance on aperture photometry, which is not optimised for crowded stellar fields where close companions cannot be disentangled. A PSF-fitting approach is implemented in the PLATO pipeline to overcome these issues. Assessing the CTI correction efficiency with such a photometry extraction method is an important future step (work in progress). For light curves generated on board using aperture photometry, the CTI correction strategy proposed in this work cannot be applied in-flight. For these cases, a dedicated on-ground correction algorithm will be required, likely leveraging knowledge of the PSF, since pixel-level information is unavailable for stars whose photometry is computed on board.
 
In our simulation, the long-term drift of the star along the focal plane does not affect the transit depth inference. This is because the simulated light curves will have been de-trended prior to the transit depth being measured. The PLATO light curves will be de-trended using a variant of the PCD-MAP algorithm \citep{Smith_2012} optimised for PLATO. 

However, in realistic conditions, a complex coupling between CTI and drift is expected, in contrast to the idealised case of our simulations. Therefore, under actual operations, this de-trending will not be perfect and may leave residuals that could introduce bias in the transit depth measurement.

Our model for the spatial variation in trap density assumes a radial distribution with a gradual increase towards the edges of the focal plane. This assumption is primarily supported by radiation analysis for PLATO cameras, although the analysis reveals moderate large-scale departures from perfect axisymmetry. Improving this model will require introducing azimuthal components, which remain to be defined. In principle, analysing CTI-induced trails from stellar images at different CCD positions could refine the model and enhance correction accuracy; this approach warrants a dedicated feasibility study.

The capture cross-sections (\(\sigma_{k}\)) for each species were fixed using the values reported in \citet{Prodhomme_2016}, which are consistent with those found in the literature \citep{Prodhomme_2014, Hall_2015}. The EPER method, however, does not allow for the derivation of capture cross-sections. To obtain these values, one must instead use the first pixel response (FPR) method \citep[see e.g.][]{Prodhomme_2014, Prodhomme_2014b}. The FPR method involves studying the charge loss that occurs at the leading edge of the charge block transferred through the CCD due to CTI. However, it remains an open question whether PLATO data can be used effectively to apply the FPR method, which is a topic we plan to address in the future.

We assume that the radiation dose increases slowly over time, and so calibration is planned at the start of each three-month observation period. This assumption may fail in the event of major solar activity, such as flares, which can cause rapid surges in radiation. In such cases, additional calibration using CCD images acquired immediately after the event would be necessary.

\begin{figure}
  \centering
  \includegraphics[width=\columnwidth]{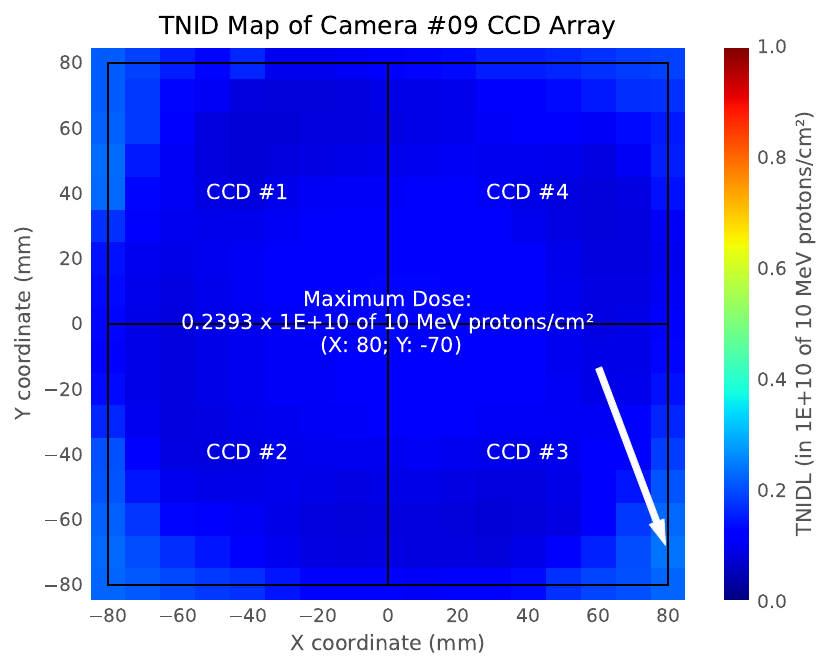}
\caption{Radiation map of Camera~9 illustrating its lower radiation exposure compared to Camera~1 in Fig.~\ref{fig:og_radiation_map}. Camera~9 is located near the centre of the optical bench, whereas Camera~1 is positioned at a corner. Adapted from \citet{ohb2024radiation}.}
  \label{fig:camera_9_radiation_map}
\end{figure}

Our validation is based entirely on simulated datasets, as it is infeasible to corroborate the results with real in-flight images. A promising path forwards is to test the correction scheme using transit light curves obtained during operations with different cameras. 
Indeed, CTI will not affect all CCDs and cameras uniformly. As shown in Fig.~\ref{fig:camera_9_radiation_map}, Camera~9 exhibits the lowest radiation exposure, in contrast to Camera~1 in Fig.~\ref{fig:og_radiation_map}. Such variations, both across cameras and among CCDs within a camera, translate into differences in trap distribution and, consequently, in the severity of CTI. Hence, comparing transit light curves obtained on cameras with very different radiation damage should in principle allow us to validate the efficiency of the CTI correction.

Addressing all these challenges, in parallel with refining our correction algorithm, is essential to ensuring PLATO meets its scientific objectives in the detection and characterisation of Earth-sized exoplanets. These challenges will be addressed in future work.

\section*{Data availability}
\label{sec:data_avail}

The CTI calibration and correction pipeline described in this paper is available on GitHub at \href{https://github.com/mishrashaunak/plato-cti-correction}{github.com/mishrashaunak/plato-cti-correction}. The underlying data generated for the simulations are available from the corresponding author upon request.

\begin{acknowledgements}
This work presents results from the European Space Agency (ESA) space mission
PLATO. The PLATO payload, the PLATO Ground Segment and PLATO data processing
are joint developments of ESA and the PLATO Mission Consortium (PMC). Funding for
the PMC is provided at national levels, in particular by countries participating in the
PLATO Multilateral Agreement (Austria, Belgium, Czech Republic, Denmark, France,
Germany, Italy, Netherlands, Portugal, Spain, Sweden, Switzerland, Norway, and United
Kingdom) and institutions from Brazil. Members of the PLATO Consortium can be found
at \url{https://platomission.com/}. The ESA PLATO mission website is
\url{https://www.cosmos.esa.int/plato}. We thank the teams working for PLATO for all their work.
This work has benefited from financial support by Centre National d'Etudes Spatiales (CNES) in the framework of its contribution to the PLATO mission. We acknowledge P. Verhoeve and T. Prod'homme for the very useful discussions about CTI. We also thank Henning Wulf for providing access to the OHB Technical Report on PLATO Radiation Analysis. S.M.'s work has received support from France~2030 through the project named \textit{Académie Spatiale d'Île-de-France} (\url{https://academiespatiale.fr/}) managed by the National Research Agency under the reference ANR-23-CMAS-0041.
The results presented in this paper were possible with the help of the Scipy \cite{2020SciPy-NMeth} and  Numpy \cite{2020NumPy-Array} libraries.
 
\end{acknowledgements}

\bibliographystyle{aa} 
\bibliography{ref}

\end{document}